# An MSX infrared analysis of the superbubble around the giant HII region NGC3603

WANG Jia & CHEN Yang[*]

*Department of Astronomy, Nanjing University, Nanjing 210093, China*

Using the MSX mid-infrared observations, we reveal a 100pc-scale superbubble surrounding the giant HII region NGC3603. We suggest that the diffuse surrounding infrared emission in bands A, C, and D is dominated by that of PAH and the emission in band E is dominated by that of dust grains. The fitted dust temperature is consistent with heating by the central cluster's UV photons. The derived gas-to-dust mass ratio for the bubble shell is of order $10^2$.

**superbubble, HII region, cluster, PAH, dust**

NGC3603 is one of the most massive HII regions in the Milky Way, located in the Carina spiral arm (RA=$11^h$, Dec=$-61°$) at a distance of $d$ =7 kpc [1]. The dense open cluster in NGC3603 is very young ($t$ ~1 Myr) [2], harboring ~$10^3$ stars [3]. With an integrated Hα luminosity of $L$(Hα) ~$2 \times 10^{39}$ erg s$^{-1}$ and a total gas mass of $4\times10^5 M_\odot$, it qualifies as a giant HII region, with a mass comparable to extragalactic giant HII regions [4, 5]. In this report, we present a preliminary study of the diffuse infrared (IR) emission from the superbubble surrounding this giant HII region.

## 1 Data and results

The IR data of the NGC3603 region are obtained from the mid-IR (centered on 8.28, 12.13, 14.65 and 21.34 μm) imaging survey of the Galactic plane, which was carried out by the Spirit III instrument onboard the Midcourse Space Experiment (*MSX*) satellite during the initial 10-month phase of the mission (terminated on February 20, 1997). We get the image data of NGC3603 from the *MSX* Galactic Plane Survey. Figure 1 shows the large scale (at least 2°×2° or 244 pc × 244 pc) diffuse gaseous structure surrounding the central bright giant HII region. A patch of bright nebular IR emission ~24′ southwestward from the center corresponds to the HII region NGC3576 in the foreground (at a distance of ~3 kpc) [1]. A superbubble surrounding NGC 3603 can be discerned, with a shell of radius $R$ ~28′ or 60 pc extending from southwest to northeast.

To study the diffuse infrared emission surrounding the giant HII region, we remove the point sources in the field of view as well as the bright central and southwestern nebulous patches. To make the analysis tractable, we focus the diffuse IR emission in the annulus region between the two circles ($r_1$=1° and $r_2$=24′) defined in Figure 1 and the northwestern section of the luminous shell (the rectangular area in Figure 1, called "region A"). With a subtle background subtraction and *MSX*'s isophotal assumption[a], we get the fluxes of the whole shell and the northwest shell rim in Region A in *MSX*'s bands A, C, D, and E (see Table 1).

In bands A and C, there is a class of broad emission features, sometimes called "unidentified infrared emission bands" (UIBs; at 3.3, 6.2, 7.7, 8.6, 11.3, 12.7 μm). Some of these bands are widely believed to be typified by the bending and stretching modes of C=C and C-H bonds in aromatic molecules, e.g. fluorescent emission of polycyclic aromatic hydrocarbons (PAHs) [6]. However, other contenders also exist in the literature (e.g. amorphous materials with aromatic hydrocarbons [7]). Band A includes the dominant UIB features at 7.7 and 8.6 μm and band C includes the UIB features at 11.3 and 12.7 μm [8]. Band D also includes emission due to PAH, deuterium-PAH, or others [9–11]. Band E is dominated by the continuum of cool dusty objects. We suggest that the emission of bands A, C, and D is dominated by the emission of PAHs.

---

*Corresponding author (email: ygchen@nju.edu.cn)

[a] http://irsa.ipac.caltech.edu/applications/MSX/MSX/imageDescriptions.htm

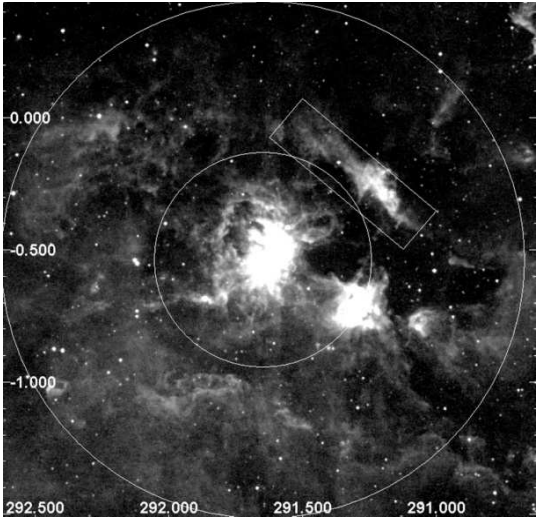

**Figure 1** NGC3603 image (at 8.28 μm).

**Table 1** Flux of annulus (super bubble) region and shell region

| Band (isophotal wavelength) (μm) | 50% peak Intensity (μm) | Flux ($10^3$Jy) (annulus) | Flux ($10^3$Jy) (shell) |
| --- | --- | --- | --- |
| A (8.28) | 6.8−10.8 | 3.72±0.77 | 0.38±0.055 |
| C (12.13) | 11.1−13.2 | 4.85±1.45 | 0.60±0.098 |
| D (14.65) | 13.5−15.9 | 3.21±0.76 | 0.39±0.058 |
| E (21.34) | 18.2−25.1 | 6.72±2.06 | 1.10±0.21 |

Using Kirchhoff's law, the dust emission flux is given by

$$F_d(\nu) = Z\nu^4 / \left(e^{h\nu/kT_d} - 1\right), \quad (1)$$

where

$$Z = 2\pi V q n_d h a^2 c^{-2} d^{-2}, \quad (2)$$

in which $V$ is the volume, $n_d$ the number density of dust grains, and $a$ the grain radius. Here, we have adopted the absorption efficient factor $Q_a = q\nu$ for grains [12], where $q$ is a coefficient. We fit the IR emission of bands A, C, D, and E with a "3PAHs+greybody" model:

$$F(\nu) = \begin{cases} U + F_d(\nu_A), & \text{band A}, \\ aU + F_d(\nu_C), & \text{band C}, \\ abU + F_d(\nu_D), & \text{band D}, \\ F_d(\nu_E), & \text{band E}, \end{cases} \quad (3)$$

where $U$ is the contribution from the PAHs in the band A flux. The PAHs' contribution in band C is scaled as $aU$, and that in band D is as $abU$. For $T_d \sim 30$ K, we have $F_d(\nu_A) \ll F_d(\nu_C) \ll F_d(\nu_D) \ll F_d(\nu_E)$ and thus the PAH emission dominates bands A, C, and D. The least-squares best-fit parameters $a, b, U, Z$, and $T_d$ are summarized in Table 2.

**Table 2** The fitting and derived parameters for dust

| Region | $a$ | $b$ | $U$ ($10^2$ Jy) | $Z$($10^{-42}$ JyHz$^{-4}$) | $T_d$ (K) | $n_d$ (cm$^{-3}$) |
| --- | --- | --- | --- | --- | --- | --- |
| Annulus | 1.3 | 6.6×10$^{-1}$ | 37.2 | 7.7 | 38.3 | 1.5×10$^{-10}$ |
| Region A | 1.6 | 6.5×10$^{-1}$ | 3.7 | 2.6 | 36.7 | 3.0×10$^{-9}$ |

## 2 Discussion and conclusions

### 2.1 The dynamics of the superbubble

We suggest that the superbubble is created by the superwind from the central star cluster of NGC3603. From the canonical wind-bubble evolution law [13]: $R = 0.76(L_w/\rho_0)^{1/5} t^{3/5}$ (where $\rho_0$ is the mass density of the ambient material and $L_w \sim 10^{39} L_{39}$ erg/s is the superwind power), we have $\rho_0 \sim 3.7 \times 10^{-23} L_{39}$ g cm$^{-3}$ if $t \sim 1$ Myr is adopted. The shell expansion velocity is about $0.6R/t \sim 35$ km/s.

### 2.2 The temperature of the dust grains

The dust around NGC3603 is warmed by the ultraviolet emission from stars in the cluster. Thus, the temperature of the dust grains is determined by [12] $T_d = T_{\text{eff}} W^{1/5}$, where $T_{\text{eff}}$ is the effective temperature and $W = \sum R_*^2/(4R^2)$ is the dilution factor, with $R_*$ the stellar radii of stars in the cluster. Using the evolutionary synthesis model GISSEL95 [14] for bolometric correction (for stellar mass between $2.5 M_\odot$ and $125 M_\odot$), we have $T_{\text{eff}} = 3.3 \times 10^4$ K. With Salpeter initial mass function, $W \sim 1.5 \times 10^{-15}$. Thus, the dust grain temperature $T_d \sim 36$ K, very close to the fitted values.

### 2.3 The gas-to-dust ratio for the bubble shell

We assume that the emission volume represented by the annulus is a sphere with radius $r_1$ excavated by a cylinder with radius $r_2$, and then this volume is $\sim 5.4 \times 10^6$ pc$^3$. The emission volume represented by the small box in Figure 1 is assumed to be a cuboid with line-of-sight length of $r_2$ and thus is $9.3 \times 10^4$ pc$^3$. Adopting $q \sim 1.45 \times 10^{-16}$ and $a = 0.15$ μm [15], we get the number densities of dust grains from eq. (2), which are also listed in Table 2.

The gas in the shell region is considered to be swept up by the radiative bubble shock (with the compression ratio $K$ of order $10^2$ [12]). The gas-to-dust ratio can be written as

$$\frac{\rho_g}{\rho_d} = \frac{K\rho_0}{n_d \rho_s 4\pi a^3/3}, \quad (4)$$

where $\rho_s$ is the mass density of solid material of the dust grains. Adopting the fitted grain number density in the shell $n_d \sim 3\times 10^{-9}\,\mathrm{cm}^{-3}$, the gas-to-dust ratio is $\sim 90\, L_{39}\, (\rho_s/1\mathrm{g\,cm}^{-3})^{-1}$. This seems similar to the usual value of 100−200.

*We thank Y.-H. Chu for critical reading of the manuscript and Q.-S. Gu and L. Ji for valuable help and financial support by the National Basic Research Program of China (Grant No. 2009CB824800) and the National Natural Science Foundation of China (Grant Nos. 10725312 and 10673003).*